\documentclass[number,citesort,dvips]{arxstspdf}
\usepackage{dcolumn,flushend}
\usepackage{graphicx}

\volume{25}
\issue{1}
\pubyear{2010}
\firstpage{22}
\lastpage{40}
\doi{10.1214/09-STS316}

\makeatletter
\def\independent{\perp\!\!\!\!\perp}
\newcolumntype{d}[1]{D{.}{.}{#1}}
\makeatother

\begin{document}
\begin{frontmatter}

\title{Assumptions of IV Methods for Observational Epidemiology}
\runtitle{Assumptions of IV Methods}

\begin{aug}
\author[a]{\fnms{Vanessa} \snm{Didelez}\corref{}\ead[label=e3]{vanessa.didelez@bristol.ac.uk}},
\author[b]{\fnms{Sha} \snm{Meng}\ead[label=e1]{sm456@leicester.ac.uk}}
\and
\author[c]{\fnms{Nuala A.} \snm{Sheehan}\ead[label=e2,text=nas11@ leicester.ac.uk]{nas11@leicester.ac.uk}}
\runauthor{V. Didelez, S. Meng and N. A. Sheehan}

\affiliation{University of Bristol, University of Leicester and
University of Leicester}

\address[a]{Vanessa Didelez is Senior Lecturer, Department of Mathematics,
University of Bristol, Bristol, UK \printead{e3}.}
\address[b]{Sha Meng is Research Fellow, Department of Health Science,
University of Leicester, Leicester, UK \printead{e1}.}
\address[c]{Nuala Sheehan is Reader, Department of Health Science,
University of Leicester, Leicester, UK \printead{e2}.}

\end{aug}

%
\begin{abstract}
Instrumental variable (IV) methods are becoming increasingly popular as they
seem to offer the only viable way to overcome the problem of unobserved
confounding in observational studies.
However, some attention has to be paid to the details, as not all such
methods target the same causal parameters and
some rely on more restrictive parametric assumptions than others.
We therefore discuss and contrast the most common IV approaches with
relevance to typical applications in observational epidemiology.
Further, we illustrate and compare the asymptotic bias of these IV estimators
when underlying assumptions are violated in a numerical study.
One of our conclusions is that all IV methods encounter problems in the
presence of effect modification by unobserved confounders.
Since this can never be ruled out for sure, we recommend that practical
applications of IV estimators be accompanied routinely by a sensitivity
analysis.
\end{abstract}

%

\end{frontmatter}

\section{Introduction}\label{sec:intro}

Inferring causation in observational studies is
problematic, as observed associations can often be due to other
than causal explanations, confounding being of special concern.
Randomized controlled trials (RCTs), rendering all other explanations
unlikely by design, are the accepted standard approach to
causal inference.
However, we are here interested in epidemiological applications where
it is not always possible nor desirable to carry out RCTs.
For example, it would be unethical or impractical to randomly allocate
individuals to exposures such as smoking, alcohol consumption, and complex
nutritional or exercise regimes. Furthermore, the cohort of a trial
might not be representative of the target population for which health
interventions are required
\cite{DaveySmithEbrahim2003,Lawloral2008a}.
The standard approach to causal inference from observational data
is to assume that there is no unobserved confounding, that is, that
a sufficient set of covariates has been measured. This is often
implausible and
has produced misleading results in the past, for example, regarding
the effects of hormone replacement therapy
\cite{LawlorDaveySmith2006,HRTRCT2002}.

Methods exploiting instrumental variables provide an alternative solution.
Suppose we are interested in the causal effect of some
exposure (e.g., cholesterol)
on disease (e.g., coronary heart disease), and
believe that important confounding
factors are likely but unobservable, perhaps because they are
not fully understood.
Loosely speaking, an instrumental variable (IV) is a third (observable)
variable that is
predictive of exposure, but has no direct effect on the disease and is
independent
of the unobserved confounders.
In general, it is difficult to find a variable that can
be justified as a suitable IV for any particular problem.
For randomized trials with partial compliance, where the effect of the
actual treatment taken is of interest,
the natural IV is the randomization to treatment \cite{Greenland2000};
but, of course, this is not an option when considering exposures that cannot
be randomized as mentioned earlier.
Examples in epidemiological contexts are the physician's
prescription preference as an IV to assess drug effects \cite{Brookhart2007,Rassenal2009c}, cigarette price to assess the effects
of smoking
\cite{LeighSchembri2004}
or genetic variants that are associated with exposures of interest
\cite{DaveySmithEbrahim2003,Katan1986,Lawloral2008a}.
The latter has become known as \textit{Mendelian randomization} and,
due to the fact that it is currently generating a lot of interest
in the epidemiological literature, will serve as illustration throughout
(see Section~\ref{sec:MR}).

Relying only on their defining properties, IVs
can\break be used to test for or bound the causal effect
\cite{Angristal1996,BalkePearl1994,DidelezSheehan2007b,Greenland2000,HernanRobins2006,Robins1994}.
However, identification and hence
point estimates of the causal effect are only obtainable
under additional parametric and distributional assumptions.
Linear structural equation models, popular in the econometrics literature
\cite{Wooldridge1,Zohoori1}, are a well-studied model class that allows
identification.
Generalizations to nonlinear structural equations\break based on
log-linear or probit modeling, for example
\cite{Mullahy1997,WindmeijerSantosSilva1997},
are also available (see overview \cite{ClarkeWindmeijer2009}).
Inspired by the simplicity of the linear case, where the IV estimator is
given as the ratio of the coefficients from the regressions of outcome
on IV and
exposure on IV, alternative methods have been put forward replacing
these two
linear regressions by nonlinear ones.
One such example which is popular in Mendelian randomization studies with
binary outcomes is what we will call the
``Wald-type'' estimator.
This combines odds ratios or risk ratios for the
genotype-outcome relationship
with the mean difference in exposure given the genotype
\cite{Casasal2005,Casasal2006,DaveySmithEbrahim2003,Keavneyal2006,Lawloral2008b,Thompsonal2005}.

An important consideration when using IV methods is the target of
inference, that is, the precise definition of the causal parameter of interest.
In our experience, epidemiologists are mostly interested in the
population causal
effect, that is, a comparative measure of subjecting everyone in a
given population
to exposure as opposed to no exposure, as would ideally be obtained in
an RCT.
However, some prominent IV methods target causal effects within specific
subgroups. These are the effect of treatment on the {compliers}
\cite{Angristal1996,ImbensAngrist1994},
or the effect of treatment on the {treated}
\cite{HernanRobins2006,Robins1989,RobinsRotnitzky2004,VansteelandtGoetghebeur2003}.
The complier causal effect is motivated by RCTs with partial compliance and
contrasts the effect of treatment versus nontreatment for those individuals
who follow their assignment whatever it is.
In our view, the interpretation of this causal parameter is very much
bound to
the randomization scenario and we will therefore not consider it any further.
The effect of ``treatment on the treated'' can be translated as the
``effect of exposure on the exposed'' in an epidemiological context and
describes the
effect of preventing those who would normally be exposed from becoming exposed.
This particular subgroup effect is explicitly modeled by
structural mean models (SMMs) \cite{HernanRobins2006}.

In this paper we compare the above approaches with regard to their use in
observational epidemiology and focus on issues that have recently arisen,
for example, in Mendelian randomization applications, to make the
discussion concrete.
We formally consider the targeted causal parameters and the underlying modeling
assumptions of IV methods.
We argue that their assumptions should be made explicit
so that those most plausible for a given problem can
be chosen. As models are never expected to be exactly true in practice,
we complement the theoretical comparison by a numerical study of
the possible bias under violations of the assumptions.
The outline of the paper is as follows.
In Section~\ref{sec:MR} we begin by presenting the basic idea of IVs
with the example of
Mendelian randomization as recently applied to investigate the effects of
alcohol consumption.
We then introduce the main concepts of causal inference in
Section~\ref{sec:causal}, a central issue being the different notions of
causal effect parameters. Section~\ref{IVcore} gives the core conditions
characterizing an instrumental variable.
In Section~\ref{sec:est} we present the IV models that we will consider,
and provide general indications of how they interrelate.
Section~\ref{sec:num} investigates the performance in terms of relative
asymptotic bias of these methods in a numerical study where we
focus on the particular case where all variables are binary in order to
facilitate exact evaluation of the relevant quantities.
We conclude with a discussion of the implications, both for epidemiological
applications and more generally.

\section{Using a Genetic Variant as an IV}\label{sec:MR}

We will relate to Mendelian randomization throughout the paper as a concrete
application of an IV approach in observational epidemiology and outline the
basic idea here using an example taken from Chen et al.  \cite{Chenal2008}.
Further details, including history and
nomenclature, are provided in a recent review \cite{Davey2007}.

Alcohol consumption has been found in observational studies to have a positive
effect on coronary heart disease (CHD) and negative
effects on liver cirrhosis, some cancers and mental health problems.
These findings, however, are strongly suspected to be confounded by
factors like diet, lifestyle and socioeconomic factors.
Thus, in order to inform public health recommendations on alcohol
intake, for
example, it is important to verify which, if any, of these observed associations
is in fact causal for the relevant health outcome.

The connection between the ALDH2 gene and alcohol consumption is well
established
and understood \cite{Bosron1986,Enomoto1991,George4,Yoshida1984}.
The ALDH2*2 variant is associated with an accumulation of acetaldehyde and
hence with unpleasant symptoms after drinking alcohol.
Carriers of this variant tend to limit their alcohol consumption regardless
of their other lifestyle behaviors.
Since genes are randomly assigned during meiosis, ALDH2*2 carriers should
not differ systematically from carriers of the ALDH2*1 allele in
any other respect.
In particular, there should be no association between the variant and the
unobserved confounders of the various relationships between alcohol
consumption and above health outcomes.
The plausibility of this assumption is strengthened by the fact that there
is no evidence of ALDH2 association with typical known epidemiological
confounders such as age, smoking, BMI, cholesterol, etc.
\cite{DaveySmithal2007}.
The possibility that ALDH2 affects the particular disease of interest
by any
route other than through alcohol consumption can also be excluded from the
known functionality of the gene.
Thus, for any specific disease, we should observe that
there are more *1*1 and *1*2 than *2*2 genotypes among the
affected individuals
if alcohol consumption is really causal for that disease.
The meta-analysis by Chen et al. \cite{Chenal2008},
based mainly on studies in Japanese populations,
shows that blood pressure and risk of hypertension is higher for
*1*1 than for *2*2 homozygotes, and is also higher for heterozygotes
(*1*2) than for the *2*2 homozygotes.
As the heterozygotes tend to be moderate drinkers due to less pronounced
adverse symptoms,
the study concludes that even moderate alcohol consumption is
``harmful'' for blood pressure.

The example shows how ALDH2 can be used as an IV to provide evidence
for a causal effect of the exposure by establishing that the disease
and the IV are associated:
the risks of high blood pressure and hypertension are significantly
different between the different genotypes.
As ALDH2 is assumed to have no direct effect on blood pressure or
hypertension other than through alcohol consumption,
the observed associations must be due to an effect of
alcohol consumption on blood pressure and hypertension.
Since the above assumptions define an IV, this reasoning only holds if
we can
be fairly confident that ALDH2 is a valid IV.
Hence, only well-understood genotypes can be used as IVs.
Note, this does not yet provide a point estimate of the causal effect
of alcohol
consumption on hypertension: it is merely evidence that there is such
an effect.

The number of applications of Mendelian randomization is growing rapidly
\cite{Casasal2005,DaveySmithal2005,George5,DaveySmithal2005c,George4,Lewis2,Minellial2004,Thompsonal2005};
a brief overview of some recent studies is given
in Sheehan et al. \cite{Sheehanal2008}.
Note that even when a genetic variant can be found that is associated
with the
exposure of interest, it does not automatically qualify as an IV.
Problems could occur when there are different
subpopulations with different allele frequencies and different prevalences
of disease, for instance, \cite{CardonPalmer2003}.
Finding a suitable genetic instrument is thus a challenge as discussed in
detail in several papers
\cite{DaveySmithEbrahim2003,DidelezSheehan2007b,DidelezSheehan2007,Lawloral2008a,Nitschal2006,ThomasConti2004}.

\section{Causal Inference}\label{sec:causal}

Epidemiologists are concerned with identifying the causal effect of an
exposure $X$ on a disease $Y$, typically with the view to
informing public health interventions.
We therefore regard causal inference to be about the effect of intervening
in, or manipulating, a given system as is implicit in many approaches to
causal inference
\cite{Dawid2002,DidelezSheehan2007,Hernan2004,Lauritzen2000,Pearl1995,Robins1989,Rubin1974,Rubin1978}.

It is useful to introduce notation to represent intervention.
Pearl \cite{Pearl2000} uses the \textit{do operator} to distinguish between
conditioning on an intervention in $X$, $P(Y|\operatorname{do}(X=x))$, and the
usual conditioning on observing $X$, $P(Y|X=x)$.
The former reflects how the distribution of $Y$ should be modified
when $X$ has been forced to the value $x$ by some external intervention,
whereas the latter reflects how the distribution of $Y$ should be
modified when $X=x$ is simply observed.
The different conditions, observation versus intervention, reflect
the common wisdom \textit{correlation is not causation}. Note that we
often write
$\operatorname{do}(x)$ for $\operatorname{do}(X=x)$.

Another formal approach is based on
counterfactual (potential outcome) variables
\cite{Hernan2004,Rubin1974,Rubin1978}.
Here $Y(x_1)$ denotes the value that the outcome $Y$ would
have if the variable $X$ were set to the value $x_1$, whereas $Y(x_2)$
is the
outcome \textit{if} the same variable $X$ were set to the value $x_2$.
The variables $Y(x_1)$ and $Y(x_2)$ are counterfactual because they can never
both be observed together, so when one is fact, the other one is, of necessity,
contrary to fact.
The notion of intervention also underlies the counterfactual
approach
\cite{Hernan2004,Robins2,Rubin1974,Rubin1978}.
Both approaches define a formal language for causality and
provide specific mathematical notation for representing interventions that
we might be interested in.
Hence, they force us to be clear and explicit about
any assumptions underlying a given method of causal inference.

\subsection{Causal Parameters}\label{MCE}

Causal effect parameters are typically functions of the distribution of $Y$
under different interventions in~$X$.
The most popular is the average causal effect (ACE)
defined as the expected difference in $Y$ under two different settings
of $X$:
\begin{eqnarray}
\operatorname{ACE}(x_1, x_2):=
E (Y|\operatorname{do}(x_2))-E(Y|\operatorname{do}(x_1)),
\nonumber
\end{eqnarray}
where $x_1$ is typically some baseline value.
The ACE is a natural choice of causal parameter when the effect of $X$ is
suspected to be linear on $Y$.
When $Y$ is nonnegative or binary, in contrast, it is more common to
use a multiplicative
measure like the causal relative risk (CRR) defined as
%
\begin{equation}
\operatorname{CRR}(x_1, x_2):=\frac{E(Y|\operatorname{do}(x_2))}{E(Y|\operatorname{do}(x_1))},\label{def_crr}
\end{equation}
or, for binary $Y$, the causal odds ratio (COR) given by
\[
\operatorname{COR}(x_1,x_2):=\frac{P(Y=1|\operatorname{do}(x_2))P(Y=0|\operatorname{do}(x_1))}
{P(Y=0|\operatorname{do}(x_2))P(Y=1|\operatorname{do}(x_1))}.
\]
Note that the odds ratio is mainly used in case-control studies to approximate
the relative risk in the case of a rare disease.

All these causal parameters are \textit{population} parameters, that
is, they compare
setting $X=x_1$ with setting $X=x_2$ for the whole population of interest.
They are what is measured in a comparison of the active and control
groups in a controlled randomized experiment when all subjects comply
with their treatment assignment. In some situations, we may be more
interested in the causal effect within a
\textit{subset} of the population, that is, conditional on a specific
value of some
observed covariates. For example, we might want to know
the average causal effect of male alcohol consumption
on oesophageal cancer risk.
The above causal parameters can easily be adapted by conditioning on
covariates provided these are prior to exposure.
We will not consider this further in the present paper.

However, one particular causal subgroup effect, or local causal effect,
is very relevant in the epidemiological literature.
This is the effect of exposure on the exposed group
\cite{Greenland2000,Robins1989,Robins1994}, or
the effect of treatment received as it is known in the context of
clinical trials (cf. \cite{Elwood2007}, e.g.).
For example, we might be interested in the effect of reducing alcohol
consumption
for those individuals who would normally tend to have high alcohol consumption,
but not in the question of increasing alcohol consumption for those who normally
do not drink much. This does not quite correspond to conditioning
on observed covariates, as what the subjects
``would normally'' be exposed to in the future is not usually observable.
However, it can be assumed that if no intervention takes place,
alcohol consumption will remain high for those individuals with
existing high consumption. In counterfactual notation the corresponding
local causal relative risk, LCRR, for instance, is given by
%
\begin{equation}\label{lcrr}
\operatorname{LCRR}
:=\frac{E(Y(x)|X=x)}{E(Y(0)| X=x)},
\end{equation}
where $Y(x)$ is the value of the outcome if an individual's alcohol consumption
is set to be $x$ and $Y(0)$ is the counterfactual outcome if it is set to
be at a baseline level, while conditioning on $X=x$ means that the
``natural'' alcohol consumption is $x$.
Note that given $X=x$, we actually observe $Y=Y(x)$, so that the
numerator of (\ref{lcrr}) is equal to $E(Y|X=x)$.
This type of causal parameter can also be expressed with the $do$-notation,
but we need to distinguish between the ``natural'' value of exposure $X$
and the one that it is set to by intervention $\tilde X$.
When no intervention takes place, these two are identical, that is,
$X\equiv\tilde X$.
However, when an intervention takes place, it is assumed that
$\tilde X$ ``overrules'' $X$ so that $Y$ causally depends on $\tilde X$
while being still \textit{associated} with $X$ due to the fact that $X$ is
informative for the unobserved confounding that also predicts $Y$.
The above can then be translated to
%
\begin{equation}\label{lcrr2}
\operatorname{LCRR}
:=\frac{E(Y|X=x, \operatorname{do}(\tilde X=x))}{E(Y| X=x, \operatorname{do}(\tilde
X=0))}.
\end{equation}
See Robins, VanderWeele and Richardson \cite{Robinsal2006} and
Geneletti and Dawid \cite{GenelettiDawid2007} for more details on
how to interpret this local causal effect without counterfactual notation.
Local versions of the ACE and COR can easily be defined
analogously to the above LCRR.
Note that the term ``local'' causal effect in the IV literature is most commonly
used for the effect of treatment on the ``compliers'' in an RCT
\cite{Angristal1996,Greenland2000,ImbensAngrist1994},
which {we are not dealing with here and which} is different
from (\ref{lcrr2}).

One further causal parameter that is sometimes considered
is the \textit{individual} causal effect which is expressed with
potential outcomes
as $Y^i(x_2)-Y^i(x_1)$. It is the difference between
the potential outcomes for a specific individual $i$.
Assumptions under which the individual causal effect can be identified
are inherently untestable \cite{Dawid2000}, but may be justified given
specific subject matter background knowledge.

Finally, we want to emphasize that a population parameter like CRR
in (\ref{def_crr}) will be different from a conditional or local
parameter LCRR in (\ref{lcrr}) or from an individual causal effect when
the effect of exposure is different in different subgroups or individuals,
that is, under heterogeneity or effect modification.
For instance, those who naturally have a high alcohol consumption are
likely to be different in many other relevant but unobservable respects
than those who have a naturally low alcohol consumption and, therefore,
the effect of changing that level should be different in these two groups.
In particular, there may be no overall effect in the population (i.e., $\operatorname{CRR}=1$)
if negative and positive effects in subgroups (or individuals) cancel
each other out.
In such a situation, an estimator that targets the CRR will be biased for
the LCRR and vice versa.
We reiterate that the accepted gold standard RCT randomizing
individuals to either $x_1$ or $x_2$ \textit{always}
targets a population causal effect.

\subsection{Instrumental Variables}\label{IVcore}

The standard approach to estimating a causal parameter from
observational data
is to assume that a sufficient set of observed confounders is available
for which we then adjust
\cite{Dawid2002,Greenlandal1999b,Lauritzen2000,Pearl2000,Rubin1974}.
When there is reason to suspect additional unobserved confounding, the causal
effect cannot typically be obtained in this way.
In this situation, IV methods permit
a different way of performing causal inference by exploiting the additional
information provided by the instrumental variable.

Recall that we denote the exposure of interest (intermediate phenotype or
modifiable risk factor) by $X$ and the outcome (disease) by $Y$.
Furthermore, we let $G$ be the instrument (e.g., genotype in a
Mendelian randomization
study) and $U$ an unobserved
variable (or, more realistically, a set of unobserved variables) that will
represent the confounding between $X$ and $Y$.
The properties that define an IV are expressed in terms of
conditional independence statements where $A\independent B \vert C$ means
$A$ is independent of $B$ given $C$.
The core conditions are the following:
\begin{longlist}[1.]
\item[1.]$G \independent U$, that is, $G$ must be (marginally) independent
of the
confounding between $X$ and $Y$;
\item[2.]$G \not\independent X$, that is, $G$ must not be (marginally)
independent of $X$;
\item[3.]$G \independent Y \vert(X,U)$, that is, conditionally on $X$ and the
confounder $U$, the instrument and the response are independent.
\end{longlist}

These properties can, to a limited extent, be tested from the observable
data (i.e., without measurements on $U$) when $G,X,Y$ are all categorical.
This is because they impose certain inequality constraints on the joint
distribution $p(y,x,g)$ (see \cite{Pearl1995b,Pearl2000} for details).
Analogous constraints can also be obtained for situations where joint
observation of $(G,X,Y)$ is not possible, but separate observations
on $(G,X)$ and $(G,Y)$ are available from different studies
\cite{Ramsahai2007},
for instance, as is often the case for Mendelian randomization applications.
Furthermore, Ramsahai \cite{Ramsahaiphd}
develops a statistical test for violation of these inequality
constraints that properly accounts for the sampling variability in the
estimated probabilities.
When the data are categorical, these inequalities
should always be verified in order to detect ``gross'' violations of the above
core conditions.
However, it should be kept in mind that distributions $p(y,x,g,u)$
will exist which violate the core conditions but may have marginals
$p(y,x,g)$ that still satisfy these inequalities.
We are not aware of analogous inequality constraints that could be
checked when $X$ is continuous (but see
\cite{Bonet2001} for the case where instrument or outcome are continuous).
Categorizing continuous variables is not advisable, as it is
possible that the continuous variables satisfy the above core conditions,
while their discrete versions do not.
Hence, since a test of the inequalities can only falsify the core
assumptions but never confirm them, and since it cannot be carried out
when the exposure is continuous, it is crucial to always justify the
core conditions on the basis of subject matter or other relevant
background knowledge.

%
\begin{figure}[b]

\includegraphics{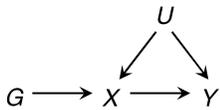}

\caption{The DAG representing the core conditions required
for $G$ to be an instrument.}\label{coreDAG}
\end{figure}

\begin{figure}

\includegraphics{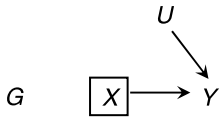}

\caption{The DAG representing the core conditions
under intervention in $X$.}\label{figureinter}
\end{figure}

A shorthand way of encoding conditional independence restrictions is via
graphical models \cite{Cowellal1999}.
The directed acyclic graph (DAG) in Figure~\ref{coreDAG} is the unique
representation of the above core conditions.
Furthermore, this graph is equivalent to a factorization of the joint
density on $(Y, X, U, G)$ in the following way:
%
\begin{equation}\label{eq1}
p(y,x,u,g)=p(y | u,x)p(x | u,g)p(u)p(g).
\end{equation}
While this describes how the variables behave ``naturally,'' we have to specify
our assumptions about how an intervention in $X$ operates on the system.
This takes the form of an additional \textit{structural} assumption
which states
that intervening in $X$ does not affect the distributions of any other factors
in (\ref{eq1})
besides the conditional distribution of $X$.
Under intervention on $X$, the joint distribution of (\ref{eq1})
thus becomes
%
\begin{eqnarray}\label{struc}
&&p(y,u,g,x | \operatorname{do}(x_0))\nonumber
\\[-8pt]\\[-8pt]
&&\quad= p(y | u, x_0)I(x=x_0)p(u)p(g),\nonumber
\end{eqnarray}
where $I(\cdot)$ is the indicator function.
The corresponding DAG in Figure~\ref{figureinter} graphically shows
the conditional independence relationships among $Y, G$ and $U$
for the core conditions and an intervention on $X$.
One immediate implication is that $G \independent Y | \operatorname{do}(X)$,
which is
also known as the {\it exclusion restriction condition} in the IV literature,
where it is typically expressed with potential outcomes as
$G\independent Y(x)$
\cite{Angristal1996,ImbensAngrist1994}.

%

\begin{figure}[b]

\includegraphics{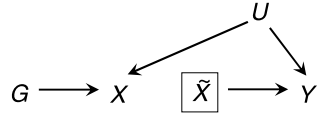}

\caption{The DAG representing the core conditions under intervention in
$\tilde X$ and ``natural'' exposure $X$.}\label{figuresmm}
\end{figure}

Taking this a step further, we can also express the IV assumptions when
the effect of exposure on the exposed individuals is of interest.
Using the notation introduced in Section~\ref{MCE}, let $X$ denote the
``natural'' exposure level, while $\tilde X$ denotes the exposure that is
set by an intervention. When there is no intervention, they are
identical and
(\ref{eq1}) is valid. Under intervention, $\tilde X$ overrules the
``natural'' $X$
with respect to the conditional distribution of $Y$ and we obtain the joint
distribution under intervention
\begin{eqnarray*}
&&p\bigl(y,u,g,X=x | \operatorname{do}(\tilde X=x_0)\bigr)
\\
&&\quad= p(y | u, x_0)p(X=x | u,g)p(u)p(g),
\end{eqnarray*}
which can again be represented graphically with a DAG as in
Figure~\ref{figuresmm}
\cite{GenelettiDawid2007,Robinsal2006}.
As before, we have the exclusion restriction
$Y\independent G|\operatorname{do}(\tilde X)$,
but we can also derive, for instance, that $Y$ is not independent of
$G$ given
$X$ and $\operatorname{do}(\tilde X)$.

%

\section{Some Common IV Models}\label{sec:est}

With the above core conditions 1--3 and structural assumption of
(\ref{struc}),
the IV can be used to test for the presence of a causal effect, or to derive
lower and upper bounds on causal effects for the case when all
variables are
categorical \cite{BalkePearl1994,Dawid2003,Manski1990,Robins1989}.
However, for general distributions of $(X,Y,G,U)$, the core conditions alone
do not necessarily allow point-identification of causal effects,
except for some extremely unusual situations \cite{Greenland2000}.

Below we present some common model restrictions, that is, additional parametric
assumptions, that enable point-identification
of causal parameters. When the causal parameter is identified, it can
be estimated consistently; in practice, small sample sizes can still
induce problems,
but we will ignore this issue here.

Our terminology is as follows. Let $\theta^*$ be the true causal
parameter of
interest, for example, the CRR
$\theta^*=E_{P^*}(Y|\operatorname{do}(x_2))/E_{P^*}(Y|\operatorname{do}(x_1))$,
where expectations are taken with respect to the true distribution $P^*$.
Restrictions are imposed in the form of a statistical model $\mathcal{M}$,
which is simply
a set of distributions with some common characteristics for the
random variables of interest, for example, the conditional mean of $Y$
being linear in $X$.
The model $\mathcal M$ is correctly specified if $P^*\in\mathcal{M}$.
The model $\mathcal M$ further allows point-identification
of the true causal effect parameter when $\theta^*$ is
equal to a function $\theta_\mathcal{M}(P^*_{X,Y,G})$ that only depends
on the
observational (i.e., not interventional)
distribution of the observable variables. The exact form of the function
$\theta_\mathcal{M}$ depends on the model assumptions, that is, on
$\mathcal M$.
If the model $\mathcal{M}$ is misspecified, then it does not contain
the true
distribution $P^*$ and $\theta_\mathcal{M}(P^*_{X,Y,G})$ will not necessarily
be equal to $\theta^*$, as the former relies on wrong model assumptions.
We call the causal parameters of interest, $\theta^*$, the \textit
{target} of inference,
and we call $\theta_\mathcal{M}(P^*_{X,Y,G})$ the \textit{estimand}
regardless of
whether the model is correctly specified or not.
Hence, the estimand is equal to the target
under a correct model and otherwise potentially different.
Note that the intervention distribution $P^*(Y=y|\operatorname{do}(x))$ itself
might be of
interest as a target, and, if identified, any causal parameter can be obtained
from it.

As we will see below, $\theta_\mathcal{M}$ can typically be expressed
in terms of
conditional probabilities or expectations with respect to the observational
distribution of
$X, Y$ and $G$. For practical data analysis, these have to be
replaced, for instance, by the corresponding empirical relative frequencies,
averages or regression coefficients, assuming that we have an independent
identically distributed (i.i.d.)
sample of $(X,Y,G)$; this then yields an estimator $\hat\theta_\mathcal{M}$.
We will not go
into the details of the actual construction of estimators $\hat\theta
_\mathcal{M}$
as functions of the sample but will focus on how different models
$\mathcal M$
allow point-identification and what the corresponding
estimands $\theta_\mathcal{M}(P^*_{X,Y,G})$ are.

Note that when parametric assumptions are made, the core conditions can
sometimes be weakened, for example, by requiring only that $G$ and $U$ are
uncorrelated, but we do not discuss this further here.
Also, some, but not all, of the following approaches are only defined when
$Y$ and/or $G$ are binary. This will be indicated when relevant.

\subsection{Linear IV Models}
The classical IV method was developed in the context of
linear models $\mathcal M$ which we define in more detail below, and
results in an
estimator $\hat\theta_\mathcal{M}$, given as the ratio of the
least squares slope estimators from linear regressions of $Y$ on $G$ and
of $X$ on $G$.
We will call this the \textit{linear IV average effect estimator},
and its estimand $\theta_\mathcal{M}$ is
%
\begin{equation}\label{eqlinearace_gen}
\mbox{LIVAE}:=\frac{\operatorname{Cov}(Y,G)}{\operatorname{Cov}(X,G)}.
\end{equation}
The \mbox{LIVAE}\vspace*{1.5pt} can equivalently be estimated by obtaining
predicted values $\hat X$ from the regression of
$X$ on $G$ and then by regressing $Y$ on $\hat X$.
It is therefore known as \textit{two-stage
least squares} \cite{Angristal1995,Wooldridge1}.
In the special case of binary instrument $G$, we have
%
%
\begin{equation}
\mbox{LIVAE}=\frac{E(Y|G=1)-E(Y|G=0)}{E(X|G=1)-E(X|G=0)}.\label{eqlinearace}
\end{equation}
This is analogous to the Wald method, which
was originally proposed to deal with the case of measurement errors in both
variables $X$ and $Y$ \cite{Boundal1995,Wald1940}.
As we shall now discuss, the \mbox{LIVAE} identifies either the
population, individual or local average causal effect (ACE, ICE or
LACE), depending
on the particular model assumptions.

In addition to the three IV conditions and structural assumption (\ref{struc}),
assume that the conditional expectation of
$Y$ is linear without interactions and
that all dependencies only affect the mean.
Then
%
\begin{eqnarray}\label{eqlinear}
\hspace*{20pt}E(Y|X=x, U=u)&=&E\bigl(Y| \operatorname{do}(X=x), U=u\bigr)\nonumber\\[-8pt]\\[-8pt]
&=&\beta x+h(u),\nonumber
\end{eqnarray}
where $h(u)$ is some function of $u$ only. With $\alpha=E(h(U))$, we have
%
\begin{equation}\label{eqdo}
E\bigl(Y| \operatorname{do}(X=x)\bigr)= \alpha+\beta x ,
\end{equation}
so that the ACE for a unit difference
in $X$ is equal to the model parameter $\beta$, while
the causal relative risk $\operatorname{CRR}(x_1,x_2)$ under this model is equal to
$(\alpha+\beta x_2)/(\alpha+\beta x_1)$.
It can easily be seen (cf. \hyperref[appendix]{Appendix})
that, under the above assumptions, $\beta=\operatorname{Cov}(Y,G)/\break\operatorname{Cov}(X,G)$.
Hence, the \mbox{LIVAE} identifies the ACE.
In the \hyperref[appendix]{Appendix} we show that the CRR is also identified in this linear model
and we will call the corresponding estimand LIVRR.

When $Y$ is binary, for example, assumption (\ref{eqlinear}) cannot
hold exactly, as it allows $E(Y|X,U)$ to take values outside [$0,1$].
It might still be used as a sensible approximation in practice,
especially when the range of $X$ is restricted and its effect is small.
As mentioned above, causal relative risks and odds ratios
that might be of more interest for binary $Y$ can also be identified
based on the linear model as detailed in the \hyperref[appendix]{Appendix}.

Under stronger model assumptions, such as those common in
the econometrics literature, for instance,
the \mbox{LIVAE} identifies the individual causal effect, ICE.
A structural equation model describes how the individual responses
$Y^i$ depend structurally (i.e., under manipulation) on other variables
\cite{Pearl2000,Wooldridge1}.
This can also be expressed using counterfactuals
\cite{Brookhart2007}.
A structural equation counterpart for (\ref{eqdo}) that parameterizes
the ICE is given by
%
\begin{equation}\label{SEM}
Y^i(x)=\beta_I x+\xi^i
\end{equation}
for individual $i$, where $\xi^i$ can be regarded as
a combination of $U^i$ and other (nonconfounding) factors that
determine the
outcome.
The problem of confounding by $U$ leads to $\xi$ and
$X$ not being independent, so that $\beta_I$ cannot be estimated
consistently from a regression of $Y$ on $X$, and the \mbox{LIVAE} is used instead.
For the interpretation it is important to note that model (\ref{SEM}) explicitly
assumes that the causal effect is the \textit{same} for each individual $i$,
while (\ref{eqlinear}) assumes that
manipulating $X$ has the same \textit{average} effect regardless
of the value of $U$ on the linear scale.
In fact, model (\ref{SEM}) implies (\ref{eqlinear}), but the converse
is not true (see the \hyperref[appendix]{Appendix} for details).

Each of models (\ref{eqlinear}) and (\ref{SEM}), together with the
IV assumptions, allows us to identify the effect of exposure on the
exposed, that is, the local average causal effect LACE, via the \mbox{LIVAE}
from (\ref{eqlinearace_gen}).
However, the LACE can be identified under weaker model
assumptions, namely, those of an additive \textit{structural mean model}
(cf. the \hyperref[appendix]{Appendix} or Hernan and Robins \cite{HernanRobins2006}).
Using the notation introduced in Section~\ref{MCE}, let $X$ denote the
``natural'' exposure level, while $\tilde X$ denotes the exposure that is
set by an intervention (overruling the ``natural'' $X$).
An additive SMM assumes that
%
\begin{eqnarray}\label{ad_SMM}
\hspace*{20pt}&&E(Y|X=x,G=g)\nonumber
\\[-8pt]\\[-8pt]
&&\quad{}-E\bigl(Y|X=x,G=g, \operatorname{do}(\tilde X=0)\bigr)= \beta_L x,\nonumber
\end{eqnarray}
where $X=0$ again denotes a suitable baseline value. Here, $\beta_L x$
is the effect of
reducing the exposure to this baseline value for those who under ``natural''
circumstances are exposed to $X=x$ and have $G=g$.
Note that this additive SMM
makes no explicit assumptions about individual causal effects or
the role of $U$; in fact, $U$ is allowed to modify the effect of $X$ on $Y$.
Implicitly, however, the manner in which $Y$ depends on $U$ is restricted
by the assumption that the above
difference in conditional expectations (\ref{ad_SMM}) does not depend
on $G$.
The different interpretation of the \mbox{LIVAE} in the context of linear
models and
presence of effect modification is also discussed by
Brookhart and Schneeweiss \cite{Brookhart2007}.

In summary,
we can use the \mbox{LIVAE} to estimate (i) the individual causal effect,
if we believe that the individual effect is the same for everyone on
the linear scale, or (ii) the average causal effect, if we believe that
the average effect is the same for different values of $U$, or (iii)
the local effect on the exposed, if we believe that this is the same for
different values of $G$.

\subsection{Nonlinear Wald Type Methods}\label{Wald}

As mentioned earlier, the \mbox{LIVAE} is the same as Wald's estimator which was
originally devised to deal with measurement errors \cite{Wald1940}.
In this section we consider two further methods leading
to ratio based IV estimators and which we will therefore call
\textit{Wald type estimators}
(cf. also \cite{Lawloral2008a,Minellial2004,Thompsonal2005}).

Several applications of Mendelian randomization,
typically considering a binary outcome $Y$, a continuous exposure
$X$ and a dichotomous genotype $G$, have used the following reasoning to
obtain an IV estimator for a causal effect
\cite{Casasal2005,Casasal2006,DaveySmithEbrahim2003,Keavneyal2006,Lawloral2008b}.
The na\"ive odds ratio of $Y$ given $X$, which we denote NOR, is suspected
to be confounded.
The odds ratio of $Y$ given the instrument $G$, which we denote by
$\operatorname{OR}(Y|G)$, is not confounded due to core
condition 3, and should be roughly equal to the causal odds ratio, COR,
between $X$ and $Y$ scaled by the mean difference in exposure for the
two genotypes,
$\delta=E(X|G=1)-E(X|G=0)$, that is, $\operatorname{OR}(Y|G)\approx$ COR$^{\delta}$.
Therefore, in these applications, the quantity NOR$^\delta$ is
compared with $\operatorname{OR}(Y|G)$ and, if
similar, the conclusion is drawn that there is no confounding and, hence,
that NOR $\approx$ COR.
We thus consider the following as the Wald type odds ratio estimand:
\[
\mbox{WaldOR}:=
\operatorname{OR}(Y|G)^{1/\delta}.
\]
(On the log-scale this is the ratio of log-odds difference and the mean
difference $\delta$, hence ``Wald type.'')
At first sight, this reasoning seems heuristic, and there is no model assumption
from which it can be theoretically derived.
However, by regarding the odds ratio as an approximation to the
relative risk for
rare diseases, we can motivate the above formula theoretically. The
following is
a slight generalization of the structural equation approach presented by
Mullahy \cite{Mullahy1997} and suitable not only for binary but also for
general nonnegative response $Y$
($X$ and $G$ can be continuous or discrete).
Assuming a log-linear model [and structural assumption (\ref{struc})],
%
\begin{eqnarray}\label{eqloglin}
&&\log E(Y|X=x, U=u)   \nonumber\\
&&\quad=\log E\bigl(Y|\operatorname{do}(X=x), U=u\bigr) \\
&&\quad=\gamma x+h(u),\nonumber
\end{eqnarray}
where $h(u)$ is some function of $u$ only.
It can then easily be seen that the causal relative risk for one unit difference
in $X$ is simply $\operatorname{CRR}=\exp\gamma$.
Further, we suppose that $X$ has conditional mean
%
\begin{eqnarray}
E(X|G=g, U=u)=\delta g +k(u),\label{eqloglin2}
\end{eqnarray}
where $k(u)$ is some function of $u$ only,
and, in addition, we require that the distribution of $X$ is such that
%
\begin{eqnarray}\label{eq:indep}
\bigl[X-\bigl(\delta G +k(U)\bigr)\bigr] \independent G \vert U.
\end{eqnarray}
Note that this requirement cannot
be satisfied when $X$ is binary, for instance, but is
automatically true when it has a conditional normal distribution.
It can now be shown (cf. the \hyperref[appendix]{Appendix} or Mullahy \cite{Mullahy1997})
that the CRR is identified because $\gamma$ is equal to
the ratio of the log-coefficient from a loglinear regression of
$Y$ on $G$ and the coefficient from a linear regression of $X$ on $G$.
In the
special case of a binary instrument $G$, this simplifies to
\[
\gamma= \frac{\log E(Y|G=1) - \log E(Y|G=0)}{E(X|G=1)-E(X|G=0)}.
\]
The method of estimating this via two regressions as mentioned above is
also called
two-stage quasi maximum likelihood \cite{Mullahy1997}.
We will refer to the estimand based on the right-hand side of the above as
the Wald relative risk (WaldRR), given by
\begin{eqnarray}\label{eqqq}
\operatorname{WaldRR}:=\operatorname{RR}(Y|G)^{1/\delta} \nonumber,
\end{eqnarray}
where $\operatorname{RR}(Y|G)$ is shorthand for the relative risk of $Y$ given
$G$. When $G$ is binary, $\delta$ is the mean difference in $X$,
otherwise it is $\operatorname{Cov}(X,G)/\operatorname{Var}(G)$.

The WaldRR identifies the CRR under the above combination of
log-linear model for $Y$ given $X$ and $U$, and the stated assumptions
on the conditional distribution of $X$ given $G$ and $U$.
Note that when $Y$ is nonnegative and continuous,
it is, in principle, possible (but not common) to elaborate
the assumptions further so that the
individual relative causal effect $Y^i(x_2)/Y^i(x_1)$ is identifiable; model
(\ref{eqloglin}) would then need to be reformulated as a structural
equation model analogously to the linear case earlier.

When $Y$ is binary and $P(Y=1)$ is small (``rare disease assumption''),
WaldRR and WaldOR will be approximately the same, so that in this case
we can argue that the WaldOR approximately identifies the COR under the same
model assumptions.
A different justification of WaldOR has been proposed by
\cite{Babanezhadal2009}
based on a logistic SMM and some very rough approximations, under which it
identifies the LCOR.

\subsection{Multiplicative Structural Mean Models}

We already mentioned that the \mbox{LIVAE} can be justified in an additive
SMM identifying the
causal mean difference within the exposed individuals (LACE).
Alternatively, we now consider a \textit{multiplicative} structural mean
model (MSMM)
\cite{HernanRobins2006}.
Again using the notation introduced in Section~\ref{MCE}, let $X$
denote the
``natural'' exposure level, while $\tilde X$ denotes the exposure that is
set by an intervention (overruling the ``natural'' $X$).
An MSMM parameterizes the LCRR (\ref{lcrr}) and is given by
%
\begin{equation}\label{MSMM}
\log
\biggl\{\frac{E(Y| X=x, G, \operatorname{do}(\tilde{X}=x))}{ E(Y| X=x, G, \operatorname{do}({\tilde X}=0))}\biggr\}=
\gamma_L x,
\end{equation}
where ${\tilde X}=0$ stands for a suitable baseline value as before.
Hence, $\gamma_L x$ is the log-relative risk of changing the exposure
to this
baseline for those who would normally be exposed to $X=x$, where it is
assumed that
the effect is the same within different levels of
the instrument $G$. This does not follow from the core IV conditions nor
from the structural assumption (\ref{struc}).
It means, for example,
that reducing the alcohol intake for those individuals who are heavy
drinkers has the same effect on the relative risk for hypertension regardless
of their ALDH2 genotype. This may be unrealistic
if those who drink much despite carrying the ALDH2*2 variant are
different in relevant aspects from those who drink much and do not
carry this
allele.
An analogous assumption is made by the additive SMM (\ref{ad_SMM}) but
for the
risk difference; note that except for trivial cases
both, the assumption that $G$ does not modify the effect on the multiplicative
\textit{and} on the additive scale, cannot be true at the same time
\cite{HernanRobins2006}.
This assumption of no
heterogeneity with respect to levels of $G$ is required
so that the model has only one unknown parameter,
since we can only identify one parameter.
When baseline covariates have been measured, it is possible to
identify more complex SMMs and this assumption could be relaxed
\cite{Babanezhadal2009,Fischeral2004,HernanRobins2006}, but we
do not consider this any further here.

In general, a SMM estimator for a causal parameter is obtained by
solving estimating equations that are based on the exclusion
restriction mentioned in Section~\ref{IVcore}.
The solution typically does not have a closed form expression.
However, for the case where $X$ and $G$ are binary, an explicit solution
exists \cite{HernanRobins2006,Robins1989} (cf. also the \hyperref[appendix]{Appendix}),
yielding that $\exp(-\gamma_L)$ equals
%
\begin{equation}\label{msmm_est}
1-\frac{E(Y| G=1)-E(Y| G=0)}{E(YX| G=1) - E(YX| G=0)}.
\end{equation}
The parameter $\gamma_L$ can easily be estimated using the
corresponding empirical
frequencies or averages.
Under the multiplicative SMM, we hence obtain that the estimand
is the inverse of (\ref{msmm_est}), which we will call MSMMRR.
It identifies the LCRR under the IV core conditions and the
assumptions of an MSMM.
In order for it to also identify the \textit{population} effect CRR,
it is sufficient to assume that $X$ and $U$ cannot interact on $Y$ on the
multiplicative scale \cite{HernanRobins2006}.
This is analogous to the ``no interaction'' assumption in linear model
(\ref{eqlinear}).
In this special case we can also obtain closed formulae for the odds ratio
and risk difference \cite{HernanRobins2006,Robins1989} (cf. also
the \hyperref[appendix]{Appendix}).

%
\begin{table*}
\caption{Summary of IV model assumptions under which the various estimands
identify the targeted causal effects\break (in addition to general
IV assumptions)}\label{tab_summary}
\begin{tabular*}{\textwidth}{@{\extracolsep{4in minus 4in}}lcp{9.7cm}@{}}
\hline
\textbf{Estimand} & \textbf{Target} & \multicolumn{1}{c@{}}{\textbf{Model assumptions}}
\\
\hline
\mbox{LIVAE}  & ICE   & Constant additive individual effect [$Y^i(x)$ linear in $x$].
\\[3pt]
\mbox{LIVAE}  & ACE   & $E(Y|X=x,U=u)$ linear in $x$, no $(X,U)$-interaction on additive scale.
\\[3pt]
\mbox{LIVAE}  & LACE  & $E(Y|X=x,G=g)-E(Y|X=x,G=g,\operatorname{do}(\tilde X=0))$
linear in $x$, no $(X,G)$-interaction on additive scale.
\\[3pt]
LIVRR  & CRR   & Same as \mbox{LIVAE} for ACE.
\\[3pt]
WaldRR & CRR   & (i) $E(X|G=g,U=u)$ linear in $g$, no $(G,U)$-interaction
on additive scale, additive independent residual.  
(ii) $E(Y|X=x,U=u)$ log-linear in $x$, no $(X,U)$-interaction on
multiplicative scale.
\\[3pt]
MSMMRR & LCRR  & $\log\{E(Y|X=x,G=g)/E(Y|X=x,G=g,\operatorname{do}(\tilde X=0))\}$
linear in $x$, no $(X,G)$-interaction on multiplicative scale.
\\[3pt]
MSMMRR & CRR   & As for LCRR, and no $(X,U)$-interaction on
multiplicative scale.
\\
\hline
\end{tabular*}
\end{table*}

Logistic structural mean models have been proposed
\cite{VansteelandtGoetghebeur2003},
but these require more restrictive assumptions, and conditions enabling
consistent estimation are relatively complicated. We therefore omit
them here.
Robins and Rotnitzky \cite{RobinsRotnitzky2004} provide a detailed
discussion of the fundamental difficulty with identifiability in SMMs,
other than for the additive or multiplicative cases.

\subsection{Comparison of Assumptions}\label{sec:comp}

The estimands, their target causal effects and the conditions for identification
are summarized in Table~\ref{tab_summary}.
(WaldOR as an approximation to WaldRR is omitted.)
The following points are noteworthy:
\begin{itemize}
\item
One could say that the strongest assumptions are those underlying
the WaldRR and WaldOR, as they rely on a
specific outcome model for the distribution of $Y$ given $(X,U),$
as well as a specific exposure model for the distribution of $X$ given $(G,U)$.
Neither the linear models nor the SMMs require the latter.
\item
All IV approaches underlying point-estimation rely on some ``no-interaction''
(or homogeneity/no effect modification assumption). No interaction between
$X$ and $U$ on the linear (or log-linear) scale in the sense of model
(\ref{eqlinear})
[or model (\ref{eqloglin})] is sufficient to ensure the assumption of
no interaction between $X$ and $G$ in the additive (or multiplicative) SMMs,
models (\ref{ad_SMM}) and (\ref{MSMM}) (see the \hyperref[appendix]{Appendix}).
However, the ``no-interaction'' assumption may either be true on the
linear or
on the log-linear scale, but not both, except in trivial cases like
$Y\independent X|U$ or $Y\independent U|X$ \cite{HernanRobins2006}.
\item
In contrast to the MSMM, the linear and Wald-type models
do not require joint information on
$(X,Y,G)$; they allow identification of the causal parameter based on separate
information on the joint distribution of $(X,G)$ and of $(Y,G)$, only.
This means that an IV analysis can be performed by exploiting results,
for example, from different existing genetic studies or meta-analyses
as is {particularly relevant for} Mendelian randomization applications
\cite{Minellial2004,Thompsonal2005}.
In addition, the WaldOR is useful for case-control studies where,
under the rare disease assumption, $\delta$ can
be approximated by a control group estimate \cite{Keavneyal2006}.
\end{itemize}

\section{Numerical Illustration of\break Asymptotic Bias}\label{sec:num}

In the previous section we have given some examples of standard models
that allow point-identification of a causal parameter exploiting an IV.
In practice, such model assumptions are unlikely ever to hold exactly,
and we
should be concerned with the robustness of IV methods under violations of
such assumptions.
Therefore, in this section we illustrate the possible bias of the
above approaches for a set of concrete scenarios that would be
realistic, for instance, in a Mendelian randomization study.
We place importance on the following issues:
\begin{itemize}
\item
A sensible IV model should allow consistent estimation
at the null-hypothesis of no causal effect.
\item
A sensible IV model should also allow consistent, or at least not
seriously biased,
estimation when there is in fact no confounding, and hence
a ``na\"ive'' analysis, based
on a regression of response $Y$ on exposure $X$ without using an IV,
would be valid.
\item
A sensible IV model should also not induce more bias than such a na\"
ive approach.
\end{itemize}
We want to investigate which of the various IV methods satisfy these desiderata,
or what situations lead to the most serious violations.

Using the notation introduced at the beginning of Section~\ref
{sec:est}, we base
our comparison on the difference between the targeted causal parameter
$\theta^*$
and the estimand $\theta_\mathcal{M}$
under a given model $\mathcal M$, evaluated at the true distribution $P^*$.
More precisely, we use the relative measure
\[
\frac{
\theta_\mathcal{M}-\theta}{\theta},
\]
which is the asymptotic relative bias of any consistent estimator
$\hat\theta_\mathcal{M}$ for $\theta_\mathcal{M}$. If the model is correctly
specified, that is, $P^*\in\mathcal{M}$, and identifies the causal
parameter, then
the above is zero.
The asymptotic relative bias can be calculated exactly, using numerical
integration where
required, under a given choice of a ``true'' joint distribution
$P^*$ of $(X,Y,G,U)$ (see below).
In special cases it is even possible to express the bias explicitly
as in \cite{Brookhart2007} for the linear case.
Note that we are not considering any sampling properties of specific estimators
$\hat\theta_\mathcal{M}$ and hence are not simulating any data.

We restrict our numerical comparison to the causal relative risk,
$\theta^*= \operatorname{CRR}$, as target. We compare the linear model, with
estimand LIVRR, the log-linear Wald type approach, with estimand WaldRR
(WaldOR is always slightly more biased for CRR than
WaldRR and is therefore omitted), and the multiplicative SMM,
with estimand MSMMRR, which all identify the CRR under their
respective assumptions as detailed in Section~\ref{sec:est}.

\subsection{Full Model}\label{sec:full}

The true joint distributions $P^*$ for $(X,Y,G,U)$ that we use for the
comparison
are specified as follows.
To facilitate interpretation and to keep the number of parametric and
distributional choices limited, we consider
dichotomous observable variables $Y$, $X$ and $G$ with the following
interpretations:
%
\begin{eqnarray*}
Y&=& \cases{
1, & diseased,\cr
0, & healthy,
}
\\
X&=&\cases{
1, & exposed,\cr
0, & not exposed,
}
\end{eqnarray*}
and we label $G=1$ to denote the value of the instrument that
predisposes to $X=1$.

The dependence of $Y$ on $X$ and $U$ is given by a logistic regression.
In addition, we assume that this model is invariant with respect to intervention
on $X$, by which we mean
%
\begin{eqnarray}\label{eqy}
&&\operatorname{logit}E(Y|X=x,U=u) \nonumber \\
&&\quad=\operatorname{logit}E\bigl(Y|\operatorname{do}(X=x),U=u\bigr)\\
&&\quad =\alpha_1+\alpha_2x+\alpha_3u+\alpha_4xu. \nonumber
\end{eqnarray}
The conditional distribution of $X$ given $G$ and $U$ is also
determined by a logistic dependence:
%
\begin{eqnarray}\label{eqx}
&&\operatorname{logit}E(X|G=g, U=u)\nonumber
\\[-8pt]\\[-8pt]
&&\quad=\beta_1+\beta_2g+\beta_3u+\beta_4gu.\nonumber
\end{eqnarray}

Finally, the marginal distribution of $G$ is determined by $p_g=P(G=1)$,
which we set to 50\% throughout (all estimands are unaffected by $p_g$),
while $p(u)$ is continuous and set to have a
uniform distribution on $[0,1]$.

The true CRR can easily be calculated from the above
using (\ref{struc}) and integrating out as
%
\begin{equation}\label{crr_int}
\frac{\int\{1+\exp(-\alpha_1-\alpha_2-\alpha_3 u -\alpha_4 u)\}^{-1}
p(u)\,du}
{\int\{1+\exp(-\alpha_1-\alpha_3 u)\}^{-1}
p(u)\,du}
.\hspace*{-28pt}
\end{equation}
Note that the CRR does not depend on (\ref{eqx}), but $\theta_\mathcal{M}$ does for
the IV models considered here.

For the above true distributions $P^*$, all models from Section
\ref{sec:est} are essentially misspecified, since
none of them model a logistic dependence of $Y$ on $(X,U)$.
Exceptions are $\alpha_2=\alpha_4=0$, or for the linear and MSMM when
$\alpha_3=\alpha_4=0$. Also, note that if $\alpha_4=0$, then there is
no effect
modification by $U$ on the \textit{logistic} scale. This does not
strictly imply
no effect modification on the additive or multiplicative scales, though
departure
from these assumptions will be more extreme when $\alpha_4\not=0$.

Our choice of $P^*$ is motivated by the fact that a logistic model like
(\ref{eqy})
would be the standard model assumption for a binary outcome if the confounder(s)
$U$ \textit{could be observed}. It is noteworthy that this default
model assumption
for the case of observed confounding is not necessarily compatible with standard
IV methods for unobserved confounding.

\subsubsection{Settings of the parameters}

There are eight parameters in (\ref{eqy}) and (\ref{eqx}).
By varying these, we consider the following set of scenarios which we
regard as realistic for epidemiological studies based on Mendelian
randomization,
for example.

We choose three strengths for the causal effect:
none $(\operatorname{CRR}=1.0)$, small $(\operatorname{CRR}=1.33)$ and
large $(\operatorname{CRR}=3.03)$; this is obtained by adjusting $\alpha_2$ accordingly.
Confounding is varied by setting $\alpha_3 \in
\{0,\break 0.1,1,2\}$, while keeping $\beta_3=2$ fixed.
Interactions are investigated by varying $\beta_4,\alpha_4\in\{-1,0,1\}$,
but note that we only consider combinations where $|\alpha_4|\leq
|\alpha_3|$,
as large interactions with small main effects are commonly perceived as
unrealistic.
The remaining parameters are chosen so as to satisfy the following
criteria. The strength of the
association between $G$ and $X$ is kept constant at
a relative risk of 2.4 throughout by adjusting $\beta_2$ accordingly.
We fix the marginals $P(X=1)=0.13$ and $P(Y=1)=0.03$ by setting
$\beta_1$ and $\alpha_1$ accordingly.
These latter values, respectively, are again typical for the exposure
frequencies and rare disease
situations, as are often encountered in Mendelian randomization studies.

\subsubsection{Bounds}\label{bounds}

To further characterize the chosen scenarios,
we calculated the nonparametric bounds for the CRR (and the ACE for comparison)
\cite{BalkePearl1994,Dawid2003,Manski1990,Robins1989}
for all our settings and found that they
were always extremely wide and always included the null hypothesis of
no effect.
For those settings where $\operatorname{CRR}=3.03$, for instance, the bounds
were of the order [$0.2,30$] (and about [$-0.08,0.8$] for the ACE where the
true ACE was around 0.06).
These are the ``tightest assumption-free bounds'' \cite{BalkePearl1994},
meaning that the observable frequencies $p(y,x,g)$ alone, derived from
the above
distributions by marginalizing over $U$, do not allow us to narrow down
the causal
effects any further.
This re-emphasizes the fact that point-identification via an IV model
relies heavily on the additional parametric assumptions that have to be made.
Narrower bounds can be obtained when a stronger instrument is used,
that is, by increasing the $G$--$X$ association.
However, the relative risk of 2.4 used here is about as strong as we would
expect to see in a Mendelian randomization study.

\subsection{Numerical Results}

We now compare the asymptotic biases of the LIVRR, WaldRR and MSMMRR.
In addition, we consider the na\"ive relative risk, NRR,
obtained as $P^*(Y=1|X=1)/P^*(Y=1|X=0)$, which gives an indication of
the bias
of a standard analysis when not using an IV. In our settings, the NRR
is unbiased when
there is no confounding, but not necessarily otherwise.

\subsubsection{No causal effect}
We begin with the case where $\operatorname{CRR}=1$, which usually
constitutes the null hypothesis.
When $\alpha_4=\alpha_2=0$, no table is shown as
none of the IV models from Section~\ref{sec:est} are misspecified,
only the NRR is biased by as much as 39\%.
However, $\operatorname{CRR}=1$ can also arise when $\alpha_2$ and $\alpha_4$ are
nonzero and of
opposite signs. The relative biases for the corresponding
settings are shown in Table~\ref{tab1}.

%
\begin{table}[b]
\caption{Asymptotic relative biases when estimating CRR
for all settings with $\operatorname{CRR}=1$ and $\alpha_4\not=0$}\label{tab1}
\begin{tabular*}{\columnwidth}{@{\extracolsep{4in minus 4in}}ld{2.0}d{2.0}d{1.3}d{2.3}d{2.3}d{2.3}@{}}
\hline
&&& \multicolumn{4}{c@{}}{\textbf{Relative bias}} \\
 \cline{4-7}\\[-6pt] 
$\bolds{\alpha_3}$
& \multicolumn{1}{c}{$\bolds{\alpha_4}$}
& \multicolumn{1}{c}{$\bolds{\beta_4}$}
& \multicolumn{1}{c}{\textbf{NRR}}
& \multicolumn{1}{c}{\textbf{LIVRR}}
& \multicolumn{1}{c}{\textbf{WaldRR}}
& \multicolumn{1}{c@{}}{\textbf{MSMM}} \\
\hline
1 &1&0 &0.277& 0.105& 0.110 & 0.095\\
2 && &0.414 &0.092& 0.095 &0.075\\[3pt]
1 &-1& &0.020& -0.113 &-0.108 &-0.101\\
2 && &0.174 &-0.106 &-0.102 &-0.087\\[6pt]
1& 1&1 &0.361& 0.198& 0.213 & 0.163\\
2 && &0.545 &0.177 &0.189 & 0.125\\[3pt]
1 & -1& &0.025& -0.202& -0.187&-0.169\\
2 && &0.226 &-0.195& -0.181& -0.140\\[6pt]
1&1 &-1 &0.184& 0.006& 0.006& 0.006\\
2 && &0.272 &0.002& 0.002& 0.002\\[3pt]
1 &-1& &0.013& -0.009& -0.009& -0.009\\
2 && &0.115 &-0.006& -0.006& -0.005\\
\hline
\end{tabular*}
\end{table}

The problem we mentioned earlier, and that becomes evident here, is
that there
can be two types of scenarios where $\operatorname{CRR}=1$: either there is no causal
effect of
exposure in any
subgroup ($\alpha_2=\alpha_4=0$), or there are different causal effects
in subgroups which cancel out overall.
The latter occurs when $\alpha_2$ and $\alpha_4$ are nonzero
in such a way that the ratio of integrals in (\ref{crr_int})
happens to be one.

All IV methods exhibit some bias in these scenarios, with around 20\% relative
bias in the worst case.
We can see the following patterns in Table~\ref{tab1}.
When $\beta_4=-1$, all IV estimators {are only slightly biased,}
while the NRR can be biased by up to 27\%.
There are only two settings where all IV methods
are more biased than the na\"ive one, and these are when
$\alpha_4=-1$ and $\alpha_3=1$, and $\beta_4=0$ or 1.
For all considered settings, the MSMMRR is the least biased, and the
WaldRR is
the most biased, but the order of magnitude is generally comparable and we
would not suggest an overall ranking of the approaches based on these results
alone.

Recall that the MSMMRR does not actually target the CRR, but targets
a particular subgroup effect---the local causal relative risk of
exposure \textit{within the exposed}---instead.
The latter is typically not one when $\alpha_2\not=0$.

\subsubsection{Causal effect but no confounding}
Let us now consider those scenarios where there is no confounding
(so either $\alpha_3=\alpha_4=0$ or $\beta_3=\beta_4=0$).
No plots or tables are shown here as only the WaldRR has nonzero bias.
This is because all assumptions of the na\"{i}ve, linear and
multiplicative structural mean models are satisfied when there
is no confounding and when $X$ and $Y$ are binary.
In contrast, as noted in Section~\ref{Wald} and again in the \hyperref[appendix]{Appendix},
the assumption (\ref{eq:indep}) underlying the WaldRR
cannot be satisfied when $X$ is binary.
We observed biases for the WaldRR and WaldOR
of up to $3.2$\% and $4.5$\%, respectively, for a moderate effect
size of $\operatorname{CRR}=1.33$, and biases as large as $65$\% and $76$\%, respectively,
when $\operatorname{CRR}=3.03$.

\subsubsection{Causal effect and confounding}
We now consider those scenarios where there \textit{is} a causal
effect \textit{as well as} confounding.
Tables~\ref{tab2} and~\ref{tab3} show the
results for a small causal effect ($\operatorname{CRR}=1.33$)
and a large causal effect ($\operatorname{CRR}=3.03$), respectively.

\begin{table}
\caption{Asymptotic relative biases when estimating CRR
for all settings with $\operatorname{CRR}=1.33$}\label{tab2}
\begin{tabular*}{\columnwidth}{@{\extracolsep{4in minus 4in}}l d{2.0}d{2.0}d{1.3}d{2.3}d{2.3}d{2.3}@{}}
\hline
&&& \multicolumn{4}{c@{}}{\textbf{Relative bias}} \\
 \cline{4-7}\\[-6pt]
$\bolds{\alpha_3}$
& \multicolumn{1}{c}{$\bolds{\alpha_4}$}
& \multicolumn{1}{c}{$\bolds{\beta_4}$}
& \multicolumn{1}{c}{\textbf{NRR}}
& \multicolumn{1}{c}{\textbf{LIVRR}}
& \multicolumn{1}{c}{\textbf{WaldRR}}
& \multicolumn{1}{c@{}}{\textbf{MSMM}} \\
\hline
0.1&0&0& 0.015& 0.003& 0.036& -0.000\\
1.0 && &0.150 &0.027& 0.066& -0.001\\
2.0 && & 0.299 & 0.051& 0.097& -0.002
\\[6pt]
1.0&1&0& 0.275& 0.130& 0.206& 0.093\\
2.0 &&& 0.411& 0.141& 0.222& 0.072
\\[3pt]
1.0&-1&0& 0.020& -0.085& -0.071& -0.101\\
2.0 &&& 0.172& -0.052& -0.033& -0.088
\\[9pt]
0.1&0&1&0.019& 0.005& 0.038&-0.000\\
1.0 & &&0.195 &0.048& 0.095 & -0.002\\
2.0 & &&0.392 &0.096& 0.160 & -0.004
\\[6pt]
1.0&1&1 &0.358& 0.247 & 0.380 & 0.159\\
2.0 & &&0.541 &0.273& 0.422& 0.120
\\[3pt]
1.0&-1&1 &0.025& -0.153& -0.148 & -0.169\\
2.0 & &&0.225 &-0.100 &-0.089& -0.142
\\[9pt]
0.1&0&-1& 0.010& 0.000& 0.032& 0.000\\
1.0 & && 0.100& 0.001& 0.034& 0.000\\
2.0 & && 0.197 &0.002& 0.034& 0.000
\\[6pt]
1.0&1&-1&0.182& 0.004& 0.039& 0.005\\
2.0 & && 0.270& 0.002& 0.032& 0.002
\\[3pt]
1.0&-1&-1&0.013& -0.004& 0.027& -0.009\\
2.0 & && 0.114 &-0.000& 0.034& -0.006\\\hline
\end{tabular*}
\end{table}

First, let us compare the results for small versus large CRR.
The na\"ive relative risk (NRR) behaves similarly in both cases.
The LIVRR is more biased when the
true causal effect is large---this is plausible as the nonlinearity
of the
model is more pronounced for larger causal effects.
The WaldRR is unacceptable when $\operatorname{CRR}=3.03$: with relative biases between
40\% and
250\%, it seriously overestimates the true effect.
As its bias is either comparable to, or much larger than, the bias for the
other two IV methods when $\operatorname{CRR}=1.33$, we will not consider the WaldRR any further.
The relative bias of the MSMMRR, in turn, is similar for small
and large CRR with a maximum of 17\%.

As one might expect, the LIVRR and MSMMRR
are only slightly biased, and much less so than the NRR, whenever there
is no
$X$--$U$ interaction, $\alpha_4=0$.
More surprising is that this is also the case when $\beta_4=-1$
regardless of
the other parameter values. This is not due to less confounding,
as we can see that the na\"ive relative risk is still noticeably
biased in those settings.

All methods struggle the most when $\alpha_4\not=0$ and $\beta_4=1$---the
MSMMRR bias then reaches 17\% and the extent of the LIVRR bias can
range from
24\% for small CRR to 45\% for large CRR.

Even though there is no uniformly best method, both tables show that
the MSMMRR
is much less biased in most settings. The only cases where it is
outperformed by
the LIVRR arise when $\alpha_4=-1$. The only cases where it is outperformed
by the NRR are when additionally $\alpha_3=1$.

%
\begin{table}
\caption{Asymptotic relative biases when estimating CRR
for all settings with $\operatorname{CRR}=3.03$}\label{tab3}
\begin{tabular*}{\columnwidth}{@{\extracolsep{4in minus 4in}}l d{2.0}d{2.0}d{1.3}d{2.3}d{1.3}d{2.3}@{}}
\hline
&&& \multicolumn{4}{c@{}}{\textbf{Relative bias}}\\
 \cline{4-7}\\[-6pt] 
$\bolds{\alpha_3}$
& \multicolumn{1}{c}{$\bolds{\alpha_4}$}
& \multicolumn{1}{c}{$\bolds{\beta_4}$}
& \multicolumn{1}{c}{\textbf{NRR}}
& \multicolumn{1}{c}{\textbf{LIVRR}}
& \multicolumn{1}{c}{\textbf{WaldRR}}
& \multicolumn{1}{c@{}}{\textbf{MSMM}} \\
\hline
0.1&0&0& 0.014& 0.006& 0.671 & -0.001\\
1.0 && &0.145& 0.066& 0.870 & -0.006\\
2.0 && &0.289& 0.128 & 1.090& -0.010
\\[6pt]
1.0&1&0& 0.265 & 0.161 & 1.220 &0.084\\
2.0 &&& 0.397& 0.210& 1.410 & 0.061
\\[3pt]
1.0&-1&0& 0.020& -0.036& 0.539& -0.102\\
2.0 &&& 0.167& 0.033& 0.757 & -0.093
\\[9pt]
0.1&0&1& 0.018& 0.013& 0.695 &-0.001\\
1.0 & && 0.188 & 0.132& 1.110 &-0.010\\
2.0 & && 0.379 & 0.263 & 1.630 &-0.017
\\[6pt]
1.0&1&1 & 0.344& 0.334 & 1.950 & 0.144\\
2.0 & && 0.523 &0.447 &2.510 &0.102
\\[3pt]
1.0&-1&1 & 0.025& -0.070& 0.440 & -0.170\\
2.0 & && 0.217 &0.062& 0.858& -0.150
\\[9pt]
0.1&0&-1& 0.009& -0.001& 0.647 & -0.000\\
1.0 & && 0.096& -0.006& 0.637 & -0.000\\
2.0 & && 0.191& -0.014& 0.605 & -0.000
\\[6pt]
1.0&1&-1& 0.176 &-0.019& 0.590 & 0.005\\
2.0 & && 0.261 &-0.028& 0.570 & 0.003
\\[3pt]
1.0&-1&-1& 0.013& 0.004& 0.663 &-0.009\\
2.0 & && 0.110 &-0.002& 0.648 &-0.006\\
\hline
\end{tabular*}
\end{table}

\subsubsection{Sign of bias}
Due to our choices of the coefficients of $U$, the NRR is always
positively biased.
The IV estimators can, however, be negatively biased, especially when
$\alpha_4$
or $\beta_4$ are negative. Also, their bias does not always have the
same sign.
Therefore, we cannot say that IV methods generally over- or
underestimate the true
causal effect.

\subsubsection{Other comparisons}
We also considered the other causal
parameters, ACE and COR, as targets in our chosen scenarios using the
corresponding
estimands under the three IV models.
We got broadly similar results with the SMM approach generally
producing less biased results, except in the presence of interactions,
and the Wald approach behaving very poorly throughout even
when there is little or no confounding.

All results presented so far were for scenarios with
3\% disease frequency and 13\% exposure frequency.
We also considered scenarios with 20\% disease and/or
50\% or 85\% exposure frequencies, but do not report them in detail as the
results followed similar patterns in terms of relative
performances of the various approaches.
All IV methods show much less bias with 50\% exposure
frequency, with the WaldRR performing much
more sensibly, in particular.
The MSMM is still clearly the least biased and is not
sensitive to interaction effects when the exposure frequency is 50\%.
This might be due to the exposure distribution being more balanced, so that
conditioning on $X$ is not so informative for $U$
and, hence, the local causal
effect is not much different from the population causal effect even
when there
are strong interactions.

\subsection{Practical Implications}
In Section~\ref{sec:comp} we compared the assumptions underlying the
IV models of Section~\ref{sec:est} on theoretical grounds. The above numerical
study adds the following insights:
\begin{itemize}
\item
The linear IV approach is often not considered appropriate when the
outcome variable
is binary or nonnegative. However, we found that it performed better than
expected for binary $Y$ with relative asymptotic
bias below 20\% in all but six of the considered scenarios and with less
bias than that of the na\"ive approach in all but five scenarios. This
may be
deemed acceptable, especially given the simplicity of the linear IV estimator.
However, for the linearity assumption to be at least approximately appropriate
with binary outcomes, the range of exposure $X$ should be restricted
and the true
causal effect small.
The latter is not uncommon for epidemiological---especially
Mendelian\break
randomization---applications.
\item
Although it is clear by theory alone that the Wald type methods from Section
\ref{Wald} make very strong assumptions, we have seen here that they are
not just slightly but can be \textit{extremely} biased when these
assumptions are
violated. It is especially worrisome that this occurs for realistic scenarios,
that the bias can be worse than with the na\"ive approach and increases with
the strength of the true causal effect, and that they can be biased
even when
there is no confounding since the model for the exposure $X$ is
violated. We would therefore not recommend this
approach unless there is good reason to be confident in the model assumptions.
A small true causal effect and a balanced or approximately normal
distribution of
the exposure $X$, possibly
after suitable transformation, would support this confidence.
\item
As mentioned before, all
IV approaches, excluding the bounds, make an assumption of
no-interaction or no
effect modification by the
unobserved confounder $U$ either on the additive or multiplicative
scale. The
results show that violation of this assumption
indeed seriously increases the bias of all
IV methods and can lead to bias even at the null hypothesis of no
causal effect.
In practice, this assumption is difficult to asses or justify, as
it involves the unobserved confounders which might include factors that
are poorly understood.
\item
As far as the relative bias is concerned, the MSMM approach seems
the most recommendable for situations similar to those of
Section~\ref{sec:full}, especially for binary outcomes.
However, other properties are relevant for practical
application, most important being the efficiency of the estimators.
As our numerical
study only considers a specific set of scenarios, it is also not
possible to say
whether the MSMM performs equally well in very different situations. We
therefore
recommend that further comparison and sensitivity analyses are carried
out for
any specific application.
\end{itemize}

\section{Conclusion and Discussion}\label{conclusion}

Our theoretical comparison of different IV methods was motivated by the
need for such methods in observational epidemiology, with
Mendelian randomization applications providing an example that
has generated a lot of recent interest.
The core conditions 1--3 plus the structural assumption (\ref{struc})
are sufficient for testing for a causal effect of exposure on disease,
but, as emphasized here, the identification of a causal effect
has to rely on \textit{additional} model assumptions which, if
inappropriate, can induce bias as illustrated in our numerical study.
The need for a comparison of IV methods is also highlighted by the
results of a
recent study which concluded that there were very few differences
between IV approaches because they yielded similar
results on particular data sets \cite{Rassenal2009b,Rassenal2009a}.
Our results do not support this point of view and show that
any model assumptions have to be justified carefully.

The main points to be made from our comparison are that the
different IV approaches target different parameters, where we are not
referring to the difference between a risk difference and risk ratio,
for instance, but the difference between an individual, population
or local causal effect.
In the case of the latter, the SMM approach (additive or multiplicative)
makes the weakest assumptions, as it does not require a model for the exposure
$X$ given the instrument $G$, and it only assumes (log-)linearity of the
effect within the exposed individuals.
Under stronger assumptions, essentially if
$U$ and $X$ do not interact on $Y$ on the relevant scale, the local
causal effect is equal to the population causal effect.
However, the multiplicative SMM requires joint data on the observable
variables which
may not always be available from existing studies.
For the linear model it has also been noted by
\cite{Brookhart2007}
that the traditional ratio estimator \mbox{LIVAE}
has to be given a different interpretation in the presence
of effect modification.
The Wald type estimator for the relative risk, together with the odds ratio
as an approximation to the latter, is simple and
useful for meta-analyses but makes very specific assumptions about all
conditional distributions, especially that of the exposure, and also
requires the absence of interactions on the multiplicative scale.

Our bias calculations are of course only valid for the particular model and
scenarios we chose to consider, but we believe they
still raise serious issues.
Not surprisingly, all estimators encounter difficulties in estimating the
population effect in scenarios where the \mbox{exposure} has different effects
within levels of the unobserved confounder.
Maybe more surprising are the particularly poor performances of the Wald
relative risk and odds ratio---especially in the absence of confounding.
This is supported by a recent study on odds ratio estimators
which also found that the WaldOR was often outperformed by other
approaches \cite{Babanezhadal2009}.
However, we did not find that it did ``especially well'' at the
causal null hypothesis, as reported there,
when there were interactions in the model for the outcome $Y$.
An obvious implication for practical applications of IV
methods is that the plausibility of such interactions, on the chosen
effect scale, should be explicitly addressed.
If such interactions are judged to be likely on the multiplicative scale,
then the MSMM estimator is closer to the local effect
and the Wald relative risk is likely to be seriously biased.
Also, one has to keep in mind that such interactions can induce bias
of all IV methods even at the null hypothesis of no causal effect,
though one might
hope that such exact cancellations of subgroup effects are rare.
It might be argued that, in practice, important effect modifiers will be
known and observed as additional covariates, so that once these are taken
into account, only negligible interactions with the unobserved
confounders remain,
but by definition this cannot be
verified empirically. Note that any justification for the absence of
effect modification
has to take the chosen measurement scale into account.
Due to the increased bias we have seen in our numerical study,
we would therefore recommend that practical applications of IV methods
be complemented by some sensitivity analyses, especially with regard to
such interactions in the model for the outcome $Y$.
Moreover, we would advise that these considerations are also valid for
continuous outcomes which are often analyzed unquestioned with linear
no-interaction
models.

The particularly restrictive assumptions underlying the WaldRR (and WaldOR)
raise serious concern about how to handle situations where we
do not have joint information on all the relevant variables,
such as in most meta-analyses, rendering the multiplicative SMM
estimator inapplicable. The linear IV estimator could, in principle,
be applied, as it too does not require joint data and is not as badly
biased, but for binary disease outcomes, risk differences are rarely
reported. In most applications the exposure is
continuous and robustness of the nonlinear Wald estimators to
violations in those cases remains to be investigated.
It certainly does not seem advisable to dichotomize a continuous exposure.\looseness=1

We have only considered the asymptotic bias of the various estimators.
In practice, their efficiency will also be of major concern.
It is well known that IV estimators have larger variance than the na\"ive
estimators when there is no unobserved confounding.
The variance, unlike the bias, very much depends on the strength of the
instrument, but when there is strong confounding, it is impossible to
find a strong instrument
\cite{Boundal1995,Martensal2006}.
The SMM estimators, derived from estimating equations, can be made
semi-parametrically efficient by choosing appropriate\break weights in these
equations \cite{Robins1994}. Some methods for improving the efficiency
of the
Wald type relative risk have been proposed \cite{Mullahy1997}.
Further comparisons of properties and sampling behavior of
IV estimators for the special case of a binary outcome can be found in
\cite{Babanezhadal2009,ClarkeWindmeijer2009}.

Another important issue that we have not addressed here is that of
measurement error.
Theoretically, it is not a problem if the IV is affected by measurement
error, as long as this is not differential.
If the exposure is affected by measurement error, we can still use the IV
approach to test for a causal effect.
However, all the above IV
estimators are then expected to be biased, as core condition 3 is
likely to be violated when $X$ is the measured, and not the true,
exposure.
In that case, we have to make even more modeling assumptions, namely,
about the
specific measurement error process, in order to obtain valid point estimates
\cite{VansteelandtBabanezhadGoetghebeur2007}.

\begin{appendix}

\renewcommand{\theequation}{\arabic{equation}}
\setcounter{equation}{19}

\section*{Appendix}\label{appendix}

\subsection*{Justification of \mbox{LIVAE}}
We have established that the ACE is equal to the model parameter
$\beta$ in model (\ref{eqlinear}). Define\vspace*{1pt} $\tilde G=G-E(G),$ then
$E(Y\tilde G)
=\operatorname{Cov}(Y,G)$. With core condition~1 and model (\ref{eqlinear}),
\begin{eqnarray*}
E(Y\tilde G)&=& E_{G}E(Y\tilde G|G)\\
&=& E_{G} \bigl(\beta E(X\tilde G|G) +\tilde G E(h(U))\bigr)\\
&=& \beta E(X\tilde G).
\end{eqnarray*}
Hence, $\beta=\operatorname{Cov}(Y,G)/\operatorname{Cov}(X,G)$, which is the\break \mbox{LIVAE} estimand.

Risk ratios or odds ratios require estimation of the
intercept of (\ref{eqdo}) obtained as follows:
\[
\hat{\alpha} = E(Y)- \hat\beta E(X),
\nonumber
\]
where $\hat\beta=\mbox{ LIVAE}$ from above.
Hence, the CRR and COR are identified by
\begin{eqnarray}
\mbox{LIVRR}
&:=&\frac{\hat{\alpha}+\hat\beta}{\hat{\alpha}}, \nonumber\\
\mbox{LIVOR}&:=& \frac{(\hat{\alpha}+\hat\beta)(1-\hat{\alpha})}
{\hat{\alpha} (1-\hat{\alpha}-\hat\beta)}. \nonumber
\end{eqnarray}

Further, under the additive SMM (\ref{ad_SMM}) we have by simple
rearranging that
$E(Y | X,G,\operatorname{do}(\tilde X=0))=E(Y-\beta_L X | X,G)$, where we use that
$E(Y | X=\break x,G,\operatorname{do}(\tilde X=x))=E(Y | X=x,G)$.
The exclusion restriction implies
that $Y\independent G\vert\operatorname{do}(\tilde X=0)$ (cf. Figure~\ref
{figureinter}),
which induces an estimating equation to obtain $\beta_L $
based on the moment condition
$E((Y-\beta_L X)\tilde G)=0$, where $\tilde G=G-E(G)$, as before.
The solution
is again $\beta_L = \operatorname{Cov}(Y,G)/\operatorname{Cov}(X,G)$.

\subsection*{Justification of WaldRR}
In addition to the model assumptions expressed in (\ref{eqloglin}) and
(\ref{eqloglin2}),
we need (\ref{eq:indep}), that is, the random variable $\xi:=X-E(X|G,U)$
has to satisfy $\xi\independent G|U$.
This is automatically satisfied when $X$ has a normal distribution
with constant variance given $(G,U)$, or a variance that only depends
on $U$.
More generally, this is satisfied when the
model for $X$ given $(G,U)$ is a location-scale family, where only the
location parameter depends on $G,U$, for example, the class of
(noncentral) $t$-distributions; any class that restricts the support of
the distributions it contains, like the Bernoulli, will not typically satisfy
this condition, though.

Hence, by definition, we can write $X=\delta G+k(U)+\xi$.
Consider now a regression of $Y$ on $G$ alone and substitute this expression
for $X$:
\begin{eqnarray*}
&&E(Y|G=g) \\
&&\quad= E_U E_{X|G=g,U} E(Y|X,U)\\
&&\quad= E_U[ \exp\{h(U)\} E_{X|G=g,U} \exp\{\gamma X\}] \\
&&\quad= E_U \bigl[\exp\{h(U)\} \\
&&\qquad\hspace*{19pt} \cdot E_{\xi|G=g,U}\exp\bigl\{\gamma\bigl(\delta g+k(U)+\xi\bigr)\bigr\}\bigr] \\
&&\quad= \exp\{\gamma\delta g\} E_U [\exp\{h(U)+\gamma k(U)\}\\
&&\qquad\hspace*{66pt}\cdot  E_{\xi|G=g,U}\exp\{\gamma\xi\}] \\
&&\quad\stackrel{(*)}{=} \mathit{const} \cdot\exp\{\gamma\delta g\},
\end{eqnarray*}
where $(*)$ uses $\xi\independent G|U$, so that
$E_{\xi|G=g,U} \exp\{\gamma\xi\}$ is constant in $G$.
Hence, the coefficient of $G$ in a log-linear regression of $Y$ on $G$ is
$\gamma \delta$.
Furthermore, $\delta$ can be recovered from a linear regression of $X$ on
$G$, as the latter is independent of $U$.
Thus, as stated in Section~\ref{Wald},
the CRR is identified by the WaldRR.

\subsection*{Justification of MSMMRR}
Analogously to the argument for the additive SMM,
we have by simple rearranging that
%
\begin{eqnarray}\label{smm_rearr}
&&E\bigl(Y | X,G,\operatorname{do}(\tilde X=0)\bigr)\nonumber
\\[-8pt]\\[-8pt]
&&\quad=E(Y\exp(-\gamma_L X) | X,G).\nonumber
\end{eqnarray}
The exclusion restriction
$Y \independent G \vert\operatorname{do}(\tilde X=0)$ now induces
an estimating equation to obtain $\gamma_L$
based on the moment condition
$E(Y \exp(-\gamma_L X)\tilde G)=0$, where still $\tilde G=G-E(G)$.
Due to the nonlinearity of the exponential function, this does not have
a simple
closed form solution
as in the linear case, except for binary variables as shown next.

When $G$ is binary, the exclusion restriction implies that
$E(Y | G=1,\operatorname{do}(\tilde X=0))=E(Y | G=0,\operatorname{do}(\tilde X=0))$.
By averaging over $X$,
\[
E(Y \exp(-\gamma_L X) | G=1)=E(Y\exp(-\gamma_L X) | G=0).
\]
When $X$ and $Y$ are binary as well, we obtain that $E(Y \exp(-\gamma_L
X) | G)$ is equal to
$E(YX\exp(-\gamma_L)|G)-E(YX|G)+E(Y|G)$. Hence, we can rearrange
the above equality to give (\ref{msmm_est}).

Under additional assumptions, the ACE and COR are also identified in an MSMM.
First, we note that by integrating out first $G$ and then $X$ from (\ref
{smm_rearr}),
we obtain an expression for $E(Y |\operatorname{do}(\tilde X=0))$ as
\[
e^{-\gamma_L}E(Y|X=1)P(X=1)+E(Y|X=0)P(X=0).
\]
If we assume that the $Y$-$X$ relative risk is the same within
subgroups of
$U$ as in model (\ref{eqloglin}), then $\exp(\gamma_L)$ is also the
(population) CRR (cf. also
next section).
Thus, by substituting, we now obtain an expression for
$E(Y |\operatorname{do}(\tilde X=1))$ as
\[
E(Y|X=1)P(X=1)+e^{\gamma_L}E(Y|X=0)P(X=0).
\]
From these it is straightforward to obtain the estimands that identify
the ACE or
COR by replacing $\gamma_L$ by the negative log of (\ref{msmm_est}).

\subsection*{Relations Between Assumptions}
Under the IV conditions the linear model (\ref{eqlinear})
implies the additive SMM (\ref{ad_SMM}).
As $E(Y|X=x,U=u)=E(Y|\operatorname{do}(X=x),U=u)=\beta x+h(u)$,
with definition of $\tilde X$ from Section~\ref{MCE},
\begin{eqnarray*}
&&E\bigl(Y|X=x,G=g,\operatorname{do}(\tilde X=\tilde x)\bigr)
\\
&&\quad= \beta\tilde x+ E(h(U)|G=g,X=x)
\end{eqnarray*}
and, hence,
\begin{eqnarray*}
&&E\bigl(Y|X=x,G=g,\operatorname{do}(\tilde X=x)\bigr) \\
&&\quad{} -E\bigl(Y|X=x,G=g,\operatorname{do}(\tilde X=0)\bigr)
=\beta x,
\end{eqnarray*}
which is an additive SMM.

It can be shown analogously that the log-linear model (\ref{eqloglin})
implies the MSMM (\ref{MSMM}).
In each case the reverse is not true, as discussed by
Hernan and Robins \cite{HernanRobins2006} for the special case
where all variables are binary.

Further, the structural equation model (\ref{SEM}) implies model (\ref
{eqlinear})
and hence (\ref{ad_SMM}).
The former states that the potential responses of a generic individual are
given as $Y^i(x)=\beta_I x+\xi^i$, where $\xi^i$ is fixed for the individual
but not between individuals.
Hence, across the population $E(Y(x)|U=u)= \beta_I x+E(\xi|U=u)$.
Interpreting
$E(Y(x)|U=u)$ as $E(Y|\operatorname{do}(X=x),U=u)$ and using (\ref{struc}), we obtain
$E(Y|X=x,U=u)=\beta_I x+h(u)$, which is equivalent to (\ref{eqlinear}).
The reverse
is clearly not true as counterexamples are easy to construct.
\end{appendix}

\section*{Acknowledgments}
We acknowledge research support from the Medical Research Council for
a collaborative project grant addressing causal inferences using Mendelian
randomization (G0601625) for all authors and which fully supports Sha Meng.
We are grateful to Roger Harbord, Frank Windmeijer and Jamie Robins for helpful
discussions.

\end{document}